\documentclass[%
 reprint,
 amsmath,amssymb,
 aps,
]{revtex4-1}

\usepackage{graphicx, subcaption}
\usepackage{dcolumn}
\usepackage{bm}

\usepackage{hyperref}
\hypersetup{
colorlinks, 
citecolor=blue, 
linkcolor=blue, 
urlcolor=blue
}

\usepackage{etoolbox}
\renewcommand{\doibase}[1]{https://dx.doi.org/\expandafter\notblank\expandafter{#1}{#1}{}}

\usepackage{xcolor}

\usepackage{booktabs}
\usepackage{multirow}
\usepackage{tikz}

\usepackage{caption}
\def\be{\begin{equation}}
\def\ee{\end{equation}}
\def\bea{\begin{eqnarray}}
\def\eea{\end{eqnarray}}
\def\la{\langle}
\def\ra{\rangle}

\def\e{\epsilon}

\def\ha{\hat{a}}
\def\hb{\hat{b}}

\def\Na{N_{a}}
\def\Ng{N_{b}}
\def\na{\hat{n}_{a}}
\def\n2{\hat{n}_{b}}
\def\numa{n_{a}}

\newcommand{\req}[1]{Eq.~(\ref{#1})}

\newcommand{\rfig}[1]{Fig.~\ref{#1}}

\newcommand{\oscI}{$O_{a}$}
\newcommand{\oscII}{$O_{b}$}
\newcommand{\NI}{$N_{a}$}
\newcommand{\NII}{$N_{b}$}

\usepackage[columnwise,running,displaymath, mathlines]{lineno}

\begin{document}


\title{
On the early-time behavior of quantum subharmonic generation
}

\author{Yunjin Choi}
\email{yunjinc@ucr.edu}

\author{Boerge Hemmerling}%

\author{Shan-Wen Tsai}%
\author{Allen P. Mills, Jr.}%
\affiliation{
Department of Physics and Astronomy, University of California Riverside, Riverside, CA 92521 USA
}

\date{\today}

\begin{abstract}
A few years ago Avetissian {\it et al.}~\cite{Avetissian2014,Avetissian2015} 
discovered that the exponential growth rate of the stimulated annihilation photons from a singlet positronium Bose-Einstein condensate should be proportional to the square root of the positronium number density, not to the number density itself. In order to elucidate this surprising result obtained via a field-theoretical analysis, we point out that the basic physics involved is the same as that of resonant subharmonic transitions between two quantum oscillators. Using this model, we show that nonlinearities of the type discovered by Avetissian {\it et al.} are not unique to positronium and in fact will be encountered in a wide range of systems that can be modeled as nonlinearly coupled quantum oscillators.
\end{abstract}


\maketitle

\section{\label{sec:level1}Introduction}

Creating laser radiation from gamma rays by means of the stimulated annihilation of a Bose-Einstein condensate (BEC) of positronium is an open problem in the field of atomic physics, and is generally considered a daunting challenge since the required large number densities of condensed positronium are not readily available in the laboratory. Ever since Dirac \cite{Dirac1930} used a calculation of the stimulated \cite{Einstein1917} annihilation rate to find the electron-positron annihilation cross section, and for nearly a century since \cite{Varma1977,Bertolotti1979,Ramaty1982,Loeb1986,Platzman1986,Liang1988,Platzman1990,Mills2002}, researchers have assumed that the exponential growth rate $G$ of the number $N_{\gamma}$ of stimulated annihilation photons of a gas of ultracold singlet positronium (Ps) atoms would be the stimulated annihilation cross section $\sigma=2\pi(\hbar / m_e c)^2=0.936\times10^{-20}\,\textrm{cm}^2$ times the number density $n_{\textrm{Ps}}$ of the Ps atoms times the speed of light,
\be
G = \frac{1}{N_{\gamma}} \frac{d N_{\gamma}}{dt} = n_{\textrm{Ps}} \sigma c.\label{Psrate}
\ee
Surprisingly, Avetissian {\it et al.} \cite{Avetissian2014,Avetissian2015} recently discovered that the growth rate of the stimulated emission of annihilation photon pairs from a dense collection of BEC singlet Ps atoms is in fact proportional to the square root of the Ps number density,  $\sqrt{n_{\textrm{Ps}}}$. An important implication of this discovery is that the growth rate of the stimulated annihilation for a Ps BEC should be comparatively large for the relatively small values of $n_{\textrm{Ps}}$ that could be experimentally available in the near term. 
A complementary implication is that the gamma ray gain per unit length of a high density Ps BEC will not be as large as naively thought before. Here we demonstrate that this interesting nonlinear dependence on $n_{\textrm{Ps}}$ is not an isolated phenomenon peculiar to positronium, but may also occur for other systems in which energy-conserving transitions can be modeled by the conversion of $k$ initial quanta of a first oscillator \oscI\ to $l$ final state quanta of a second oscillator \oscII. 
The particular case of the stimulated annihilation of Ps is represented by $k$=1 and $l$=2.

Avetissian {\it et al.} have extended this type of system to include the generation of coherent photon-phonon radiation in an exciton BEC \cite{PhysRevB.96.045206}, and have also studied the rate of multiphoton excitation and harmonic generation in the QED vacuum \cite{Avetissian_2018,Avetissian2016}.
Examples of other processes that should exhibit some type of early time nonlinear gain behavior, depending on the values of $k$ and $l$, include parametric subharmonic frequency generators \cite{Couteau2018,Holthaus1994,Perina2013,Sun2019}, coupled asymmetric quantum wells \cite{Batista2000}, sub-harmonic generators using single atoms \cite{Kockum2017}, nonlinearly coupled micromechanical resonators \cite{Groblacher2009}, quantum parametric oscillators with trapped ions \cite{Ding2017,Gorman2014}, and radiative decay of metastable BEC of atoms \cite{Marmugi2018}. The most well known case of harmonic generation is photon up-conversion \cite{Bloembergen1959,Auzel2004}, the conversion of two photons ($k$=2) to a single photon of half the wavelength ($l$=1). It is to be noted that the peculiar early time behavior we are considering may not have been noticed in these systems if the goal was simply to generate high amplitude harmonics or subharmonics. Careful experiments to examine the turn-on behavior of these systems would be illuminating.

In what follows, we demonstrate that the stimulated emission of pairs of annihilation gamma ray photons with equal energies and opposite momenta along one particular direction from a Ps BEC can be simply modeled by two coupled quantum mechanical oscillators to yield the same dynamics as predicted by Avetissian {\it et al.} 
Our result shows that the emission behavior is reproduced and does not require a full treatment of the momentum dependence of the Ps atoms, which is considered in the full quantum field theory treatment.

In particular, the first oscillator \oscI\ may represent a BEC of singlet Ps atoms. \oscI\ then has a natural frequency $2\omega_0$ with $2\hbar\omega_0\approx 2 m_e c^2$ being the energy of a singlet Ps atom.
The initial occupation number \NI\ of \oscI\ is very large \NI $\gg 1$ and is equal to the expectation value of the number operator in the Ps BEC. \oscI\ is coupled to a final state oscillator \oscII\ with $l$ = 2, which is initially in its ground state. 
The occupation number \NII\ of \oscII\ is the number of annihilation photons corresponding to a particular pair of opposite momenta modes of the annihilation photon field. There is no need in this model to have physically separate oscillators for the two annihilation photons with frequencies $\omega_0$. The second oscillator is coupled to the first subharmonic of the first oscillator. 
In the case of Ps two-photon annihilation, the excitation rate of the driven subharmonic quantum oscillator is proportional to $\sqrt{N_\textrm{a}}$. We now outline the mathematical proof of this assertion and examine the generalization of this model to higher-order coupled oscillators. 

\section{\label{two_photon_case} Generalized coupled harmonic oscillators}

We consider a harmonic generation process in which the occupation number $N_{a}$ of a first highly excited oscillator \oscI\ decreases by $k$ while increasing the occupation number $N_{b}$ of a second oscillator \oscII\ by $l$. 
To work out the dynamics of this system, we introduce the Hamiltonian of this system in second quantized form, 
\be
\hat{H}=\e_a \hat{a}^{\dag}\hat{a}+\e_{b}\hat{b}^{\dag}\hat{b}+g (\hat{a}^k\hat{b}^{\dag l}+\hat{a}^{\dag k}\hat{b}^l),\label{Hamiltonian}
\ee
where we employ units for which $\hbar=1$ with $k$ and $l$ being integers. 
The bosonic operators $\ha, \hb$ are the annihilation operators of \oscI\ and \oscII, respectively. The operators have normalized commutation relations, $\left[ \ha, \ha^{\dag} \right]=1, \left[ \hb, \hb^{\dag}\right]=1$. The Hamiltonian represents generalized down-conversion for $k<l$, up-conversion for $k>l$, and ordinary resonant energy transfer for $k=l$. In the case of positronium, where $k=1$ and $l=2$, the coupling constant in \req{Hamiltonian} is the singlet positronium annihilation rate, 
$g = \alpha_0^5 m_e c^2/2 \hbar \approx 8 \times 10^9\, \textrm{s}^{-1}$, where $\alpha_0$ is the fine-structure constant.

We begin with an initial state at time $t$ = 0
\be
|\Psi(0)\ra=|\alpha\ra_{a}|\beta\ra_{b},\label{initial.state1}
\ee
in which the fundamental mode oscillator \oscI\ is prepared in a coherent state $|\alpha\ra_a$, 
where $\hat{a}|\alpha\ra_a=\alpha|\alpha\ra_a$ and $|\alpha\ra_a =e^{-|\alpha|^2/2}\;e^{\alpha \ha^{\dag}}|0\ra_a$. In view of the largeness of the initial occupation number, it is convenient to employ coherent states which are optimally suited to taking semiclassical limits. However, we would get the same results if we had started with a Fock state, which is an eigenstate of the number operator. Again for the specific case of Ps, at high densities $n_{\textrm{Ps}}> 10^{20}\,\textrm{cm}^{-3}$ collisions will quickly drive an initial Fock state into a coherent state \cite{Glauber1963}. In any case, the initial occupation number of \oscI\ may be approximated as  \NI\ $= \la\alpha|\na |\alpha\ra_a=\la\alpha|\hat{a}^{\dag}\ha |\alpha\ra_a=|\alpha|^2$. 
The second oscillator \oscII\ is prepared in a number state  $|\beta\ra_{b}$, where the initially prepared average number of bosons of \oscII\ is \NII $(0)=\beta$.

To investigate the dynamics of the bosonic decay process, we use the  Heisenberg representation, where the time evolution of an operator $\hat{L}$ is given by the equation $\partial\hat{L}/\partial t = i[\hat{H},\hat{L}]$. We are interested in the time dependence of the expectation value \NII\ of the occupation number operator of \oscII, $\n2=\hb^{\dag}\hb$, for which the time derivative is given by,
\bea
\frac{d\n2}{dt} &=& i\left[\hat{H},\hb^{\dag}\hb \right] = -2gl\hat{y}
\label{dndt},
\eea
where we have introduced the Hermitian operators $\hat{x}$ and $\hat{y}$,
\begin{eqnarray}
\hat{x}&=&\frac{1}{2}(\ha^{\dag k}\hb^l+\ha^k\hb^{\dag l}),\\
\hat{y}&=&\frac{1}{2i}(\ha^{\dag k}\hb^l-\ha^k\hb^{\dag l}),
\end{eqnarray}
from the definition of $\ha^{\dag k}\hb^l=\hat{x}+i\hat{y}$ for simplicity. To solve the differential equation \req{dndt}, we use the following derivatives
\bea
\frac{d\hat{y}}{dt}&=&\delta_\epsilon \hat{x}
+g\left[\ha^k\hb^{\dag l},\ha^{\dag k}\hb^l \right] ,\label{eq:dydt}\\
\frac{d\hat{x}}{dt}&=&-\delta_\epsilon \hat{y},\label{dx.eq}
\eea
where we define the resonance detuning $\delta_\epsilon=\e_a k-\e_{b}l$ of the transition from $k$ bosons of \oscI\ to $l$ bosons of \oscII. By combining the two equations \req{eq:dydt}, (\ref{dx.eq}), we investigate the characteristics of the dynamics of the number operator $\n2$ of \oscII\ for specific cases of $l$ and $k$.

\subsection{\bf Case $\mathbf{l=1}$} 
This case shows up-conversion such that $k$ bosons of \oscI\ combine to generate a single boson of \oscII\ having $k$-times higher energy. 
We then have the following differential equations:
\bea
\frac{d\hat{y}}{dt}&=&\delta_\epsilon \hat{x}+g\na ^{k-1} (k^2 \n2-\na). \label{eq:l1}
\eea
For our case of a heavily populated initial state with mean field $\alpha$, we approximate the operators $\ha,\ha^{\dag}$ by c-numbers $\alpha, \alpha^{*}$ to decouple the two fields $\ha$ and $\hb$. The solution for the average number of generated bosons, \NII\ $(t)=\la\n2(t)\ra$, can be obtained with the initial conditions $\langle\hat{y}(0)\rangle=0$ and $\langle d\hat{y}/dt\rangle|_{t=0}=-g \Na^{k}$:
\be
\Ng(t)=\frac{2g^2\Na^{k}}{C_1}
\left(1-\cos(\sqrt{C_1} t)\right)+\beta,
\ee
where we define $C_1=\delta_\epsilon^2+2kg^2\Na^{k-1}$ and are assuming $\Na \gg 1$. We see that for short times the occupation number of \oscII\ is proportional to the mean occupation number of \oscI\ to the power $k$ times $(\sqrt{C_1} t)^2$.
This is precisely what we ordinarily see for the coherent generation of the $k$th harmonic of a fundamental oscillator, no matter what the value of $k$ may be so long as $l$=1.

\subsection{\bf Case $\mathbf{l=2}$}\label{l=2}  
For any $l=2$, regardless of the specific value of $k$, we encounter a generalized version of the type of nonlinearity discovered by Avetissian {\it et al.}~\cite{Avetissian2014,Avetissian2015}. We rewrite the \req{eq:dydt} as follows 
\be
\frac{d\hat{y}}{dt}=\delta_\epsilon \hat{x}+g\Big\{\left[\ha^k,\ha^{\dag k} \right](\n2^2-\n2)-2(1+2\n2)\ha^{\dag k}\ha^k\Big\}. \label{dy.eq2}
\ee
When we introduce the semi-classical approximation for large $\Na$, we get the solution for $\Ng$ by combining \req{dndt}, (\ref{dx.eq}) and (\ref{dy.eq2}):
\be
\frac{d^2\hat{y}}{dt^2}=-\delta_\epsilon ^2\hat{y}+16g^2\Na^{k} \hat{y}
= C_2 \hat{y},
\ee
where $C_2=16g^2 \Na^{k}-\delta_{\epsilon}^2$. The generalized solution for $\hat{y}$ is 
\be
\hat{y}=\hat{A}e^{-\sqrt{C_2}t}+\hat{B}e^{\sqrt{C_2}t}.
\ee
To find the coefficient operators $\hat{A}$ and $\hat{B}$, we use the initial conditions of $\hat{x}$ and $\hat{y}$ such as $\langle\hat{y}(0)\rangle=0$ and $\langle\hat{x}(0)\rangle=0$.
We also have 
\begin{eqnarray}
\left\langle
\frac{d\hat{y}}{dt}
\right\rangle_{t=0}
=
-2 g \Na^{k}( 1 + 2\beta ). 
\end{eqnarray}
The expectation values of the coefficient operators are thus $\langle\hat{A}\rangle=-\langle\hat{B}\rangle=g\Na^{k}(1+2\beta)/\sqrt{C_2}$. When we substitute the solution of $\hat{y}$ into \req{dndt} we find   
\bea
\Ng(t)=\frac{4g^2\Na^{k}(1+2\beta)}{16g^2\Na^{k}-\delta_\epsilon^2}
\left(
e^{\sqrt{C_2}t}+e^{-\sqrt{C_2}t}-2
\right)
+
\beta\label{meanphoton2}\quad
\eea
This result gives the expectation value of the number of generated bosons as a function of time when the outgoing boson state is initially prepared in a Fock state. If the initial state of \oscII\ is prepared in any arbitrary state, we can simply replace  $\beta \rightarrow
\la\beta|\n2|\beta\ra_b$.

Note that the initial boson number $\beta$ appears in front of the exponential functions, and hence it does not affect the exponential growth rate. However, the output intensity is linearly dependent on the number of bosons in the initial state, a direct consequence of the presence of the commutator in \req{eq:dydt}. We thus expect that in the simple case of zero detuning $\delta_\epsilon=0$ the output intensity at any given time will be proportional to $1 + 2\beta$.

An interesting result comes from the appearance of $k$ only in the power of the initially prepared particle number of \oscI. In our calculation, we have simplified the real process involving a Ps BEC by ignoring the momentum dependence of the bosons in \req{Hamiltonian}. 
If we consider the process including the phase-space integration as in Ref.~\cite{Avetissian2014} 
we would find that for the case of zero detuning, $\delta_{\epsilon} = 0$, the exponential growth rate satisfies $G \propto \sqrt{\numa^k}$ from the definition of \req{Psrate}.
Moreover, when we consider the outgoing bosonic pairs in the different modes, we can check that the outgoing pairs are not coherent with one another and so the different modes are independently growing.  

For this case with $k=1$, \req{meanphoton2} above models subharmonic generation or down-conversion in which one boson with higher energy is converted into pair of bosons. 
This equation which, in the case of a Ps BEC, concerns only a single mode of the final boson state, also exhibits precisely the same time dependence as \req{meanphoton2} of Ref.~\cite{Avetissian2014}. 
The exponential growth rate results from the positronium atoms forming a BEC. At higher and higher densities, $n_\textrm{Ps}$ $>$ $10^{21} \textrm{cm}^{-3}$, the condensate fraction will decrease first due to spin exchanging two-body collisions, then due to the formation of positronium molecules via three-body collisions, and finally due to the formation of an electron-positron plasma for $n_\textrm{Ps} > 10^{22} \textrm{cm}^{-3}$ \cite{PhysRevB.7.1508,PhysRevB.34.3820}. At the highest densities the process will be the generation of a pair of bosons from two independent fermions with a gain linearly proportional to the density \cite{Sibilia1981}. 
For $k=2$, $l=2$, which is the generation of one pair of bosons from another pair of bosons, the gain scales as $N_\textrm{a}$ as one would expect. 

The above discussion shows that we can indeed seed one of the modes of a Ps BEC represented by setting $\beta = 1$ corresponding to a single external on-resonance gamma photon. In practice the seeded mode might be difficult to detect experimentally since it will only be 3 times more intense than the plethora of spontaneous modes that have $\beta = 0$. On the other hand, it could be practical to seed one Ps BEC with the large spontaneous output from a nearby collimated Ps BEC to produce a powerful directional beam of photons \cite{Avetissian2014}. However, even if the Ps density were greater than $10^{20}$\,cm$^{-3}$ such that the naive stimulated annihilation gain of \req{Psrate} would appear to be greater than the prediction of \req{meanphoton2}, the remarkable fact is that this channel, which is linear in \NI , is not present. 


\def\figuresize{6.5cm}

\begin{figure}
\begin{subfigure}[t]{\figuresize}
    \caption{For small $\Na$}
    \includegraphics[width=\figuresize]{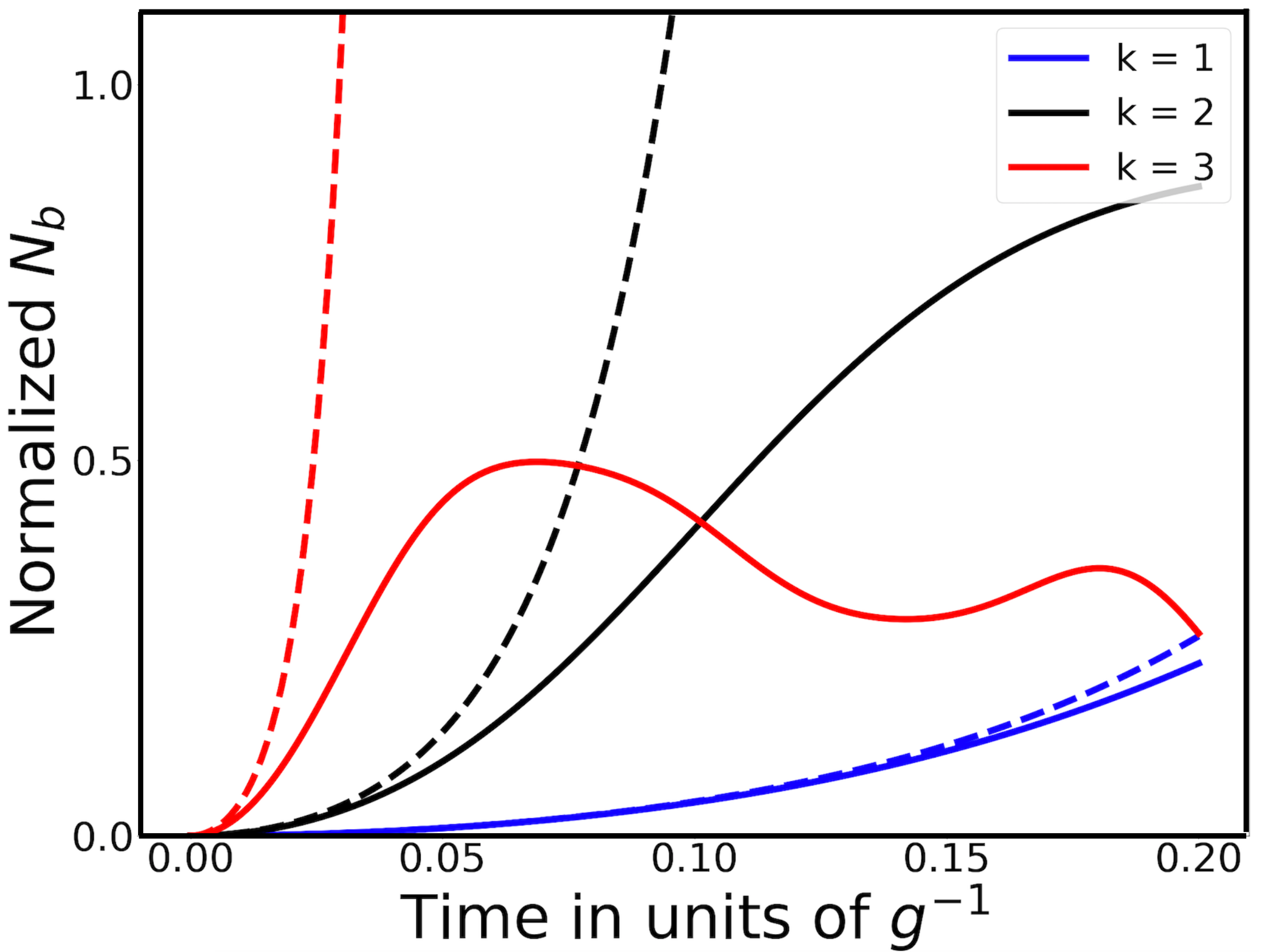}
\end{subfigure}
\begin{subfigure}[t]{\figuresize}
    \vspace*{0.5cm}
    \caption{For large $\Na$}
    \includegraphics[width=\figuresize]{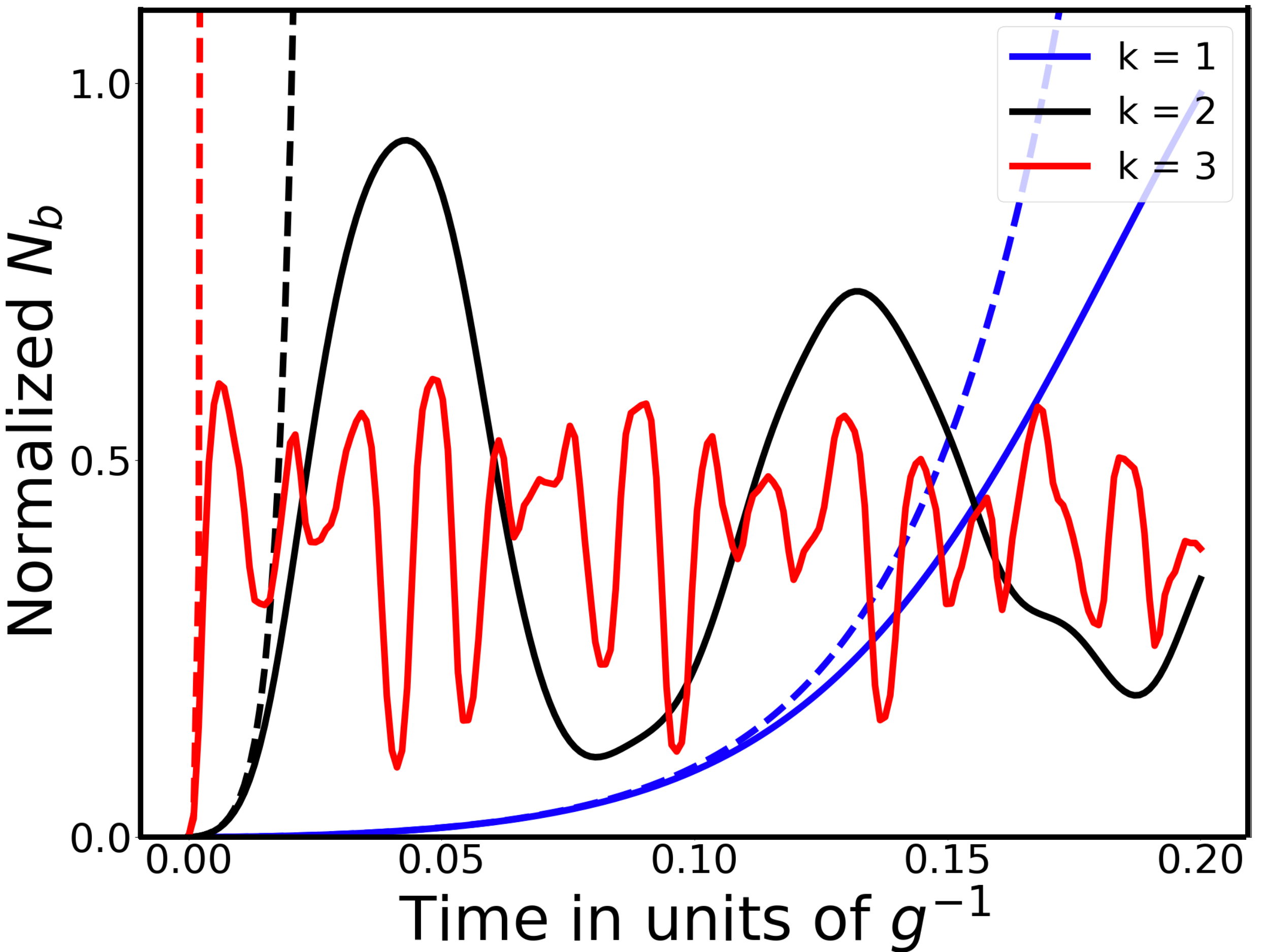}
\end{subfigure}
\caption{\label{fig:Na10} The time dependence of the expectation values of the occupation number, $\Ng(t)$, with $\Ng(0)$=0, are shown for different values of $k$ with $l=2$ when the two different initial values (a) $\Na=10$ and (b) $\Na=69$ are given. The plots show the normalized scale, $\Ng/\Na$, and compare the analytic approximations from \req{meanphoton2} (dashed lines) with the numerical results (solid lines).}
\end{figure}


\subsection{\bf Case $\mathbf{l\geqslant 3}$}
The situation for ${l\geqslant 3}$ models the generation of several lower energy bosons starting from $k$ bosons. In this case, \req{eq:dydt} contains higher powers of $\hat{b}$, and the differential equations would likely have to be solved numerically even with the semiclassical approximation. However, we may introduce the Schr\"{o}dinger picture to understand the growth rate of harmonic generation for arbitrary $l$. The time dependent quantum state is written as, 
\bea
|\Psi(t)\ra&=&e^{-i\hat{H} t}|\Psi(0)\ra\nonumber.\label{state.t}
\eea
To simplify the calculation, we take the initial state of \oscII\ to be the vacuum state, $\beta=0$. As before, we consider the initially prepared bosons to have a large population so that the mean field approximation applies. The expectation value of the boson number operator $\n2(t)$ is then given by
\bea
\Ng (t)\simeq
\frac{(l-1)!}{l}\left[\frac{t^2\bar{C}}{2!}+\frac{t^4}{4!}\Big\{\bar{C}^2 D_l-\bar{C}\delta_{\epsilon}^2\Big\} \right],\;\;\label{Schrod.pic.number}
\eea
where we consider the small time approximation $gt\ll 1$, and define the two coefficients,
\bea
\bar{C}&=&2l^3g^2\Na^k,\label{cbar}\\
D_l&=&\frac{(2l)!-2(l!)^2}{l!\;l^3}.\label{dfunction}
\eea  
Note that, $\Na^k$ in \req{cbar} is from the approximation $\prod_{i=0}^{k-1}(\Na-i)\simeq \Na^k$ for large initial boson number $\Na$.  
For $l=2$, we have seen that $\bar{C}=C_2$ and $D = 1$, so \req{Schrod.pic.number} becomes the expansion of $\frac{1}{2}(\cosh(t\sqrt{C_2})-1)$ which is same as the small time approximation of \req{meanphoton2} for the case of zero detuning, $\delta_{\epsilon}=0$. However, if either time increases sufficiently or for large $l$, $\Ng$ does not behave as the $\cosh$ function of $\Na$ and only a numerical solution is possible.
This result shows that the expectation value of the number operator initially grows as approximately $g^2t^2 \Na^k$.

We have already seen that for conversion of any number $k$ of input quanta to a single output quantum ($l$=1), the number of output quanta is proportional to the $k$th power of the number of input quanta times {$\frac{1}{2}C_1 t^2$}, as expected for ordinary frequency doubling, tripling, etc. It is conversion to a number $l$$>$1 of output quanta where we encounter exponential growth of the output with a rate proportional to the number of input quanta to the power $k/2$, $\sqrt {\Na^k}$, including the case of the coupled oscillator model for the stimulated emission of a Ps BEC when $k = 1$ and $l=2$.

In \rfig{fig:Na10} we compare the expectation values of the occupation number $\Ng$ as a function of time, predicted by the analytic approximations \req{meanphoton2}, and the numerical results from the Schr\"{o}dinger picture for both small and large values of $\Na$.
Both of the plots show that $\Ng$ begins by increasing exponentially, but the numerical result shows it converges to $2\Na/k$ after the oscillations have damped out. As $\Na$ increases, the gap between two curves (dashed and solid lines) becomes narrower, while the two curves start to separate earlier.
When $\Na$ increases, $\bar{C}$ in \req{Schrod.pic.number} increases, the fluctuation frequency becomes bigger, and the analytic approximation works only for smaller $t$.

\section{Conclusion}
 We have presented a quantum analysis of the early time behavior of sub- and superharmonic stimulated emission by modeling the conversion of $k$ initial quanta of an oscillator \oscI, to $l$ final quanta of \oscII.
We have shown that the number of final quanta $l$ determines the growth behavior of the expectation value of the number of generated bosons, while the number of initial quanta $k$ determines the power of the initial bosonic number in the equation for the growth rate.  We have demonstrated that the result of Avetissian {\it et al.} is modeled by the case $k=1$ and $l=2$. The fact that our results are applicable to any physical system which can be represented as a pair of coupled oscillators thus helps to place the Avetissian {\it et al.} discovery into context in the broader field which it has founded.

\begin{acknowledgments}
Support for this research was provided in part by the US National Science Foundation under Grants No.~PHY1839153 (Shan-Wen Tsai and Boerge Hemmerling) and PHY1505903 and PHY2011836 (Allen P.~Mills, Jr.). This work is also supported in part by the M.~Hildred Blewett Fellowship of the American Physical Society, www.aps.org (Yunjin Choi). 
\end{acknowledgments}


\bibliography{references}

\end{document}